\documentclass[11pt]{article}
\usepackage{latexsym}
\usepackage{amssymb}
\usepackage{amsmath}
\usepackage{graphicx}

\newtheorem{theorem}{Theorem}[section]
\newtheorem{definition}{Definition}[section]
\newtheorem{proposition}{Proposition}[section]
\newtheorem{assumption}{Assumption}[section]
\newtheorem{lemma}{Lemma}[section]

\newtheorem{remark}{Remark}[section]

\usepackage[usenames]{color}
\definecolor{red}{rgb}{1.0,0.0,0.0}

\definecolor{blu}{rgb}{0.0,0.0,1.0}

\def \R{\mathbb{R}}

\def \E{\mathbb{E}}

\def \P{\mathbb{P}}
\def \Q{\mathbb{Q}}

\def \Ac{{\cal A}}

\def \Cc{{\cal C}}

\def \Ec{{\cal E}}
\def \Fc{{\cal F}}
\def \Gc{{\cal G}}
\def \Hc{{\cal H}}

\def \Pc{{\cal P}}

 \def \Nc{{\cal N}}
\def \Sc{{\cal S}}

\def \ni{\noindent}

\def \eps{\varepsilon}

\def \ep{\hbox{ }\hfill$\Box$}

\long\def\symbolfootnote[#1]#2{\begingroup%
\def\thefootnote{\fnsymbol{footnote}}\footnote[#1]{#2}\endgroup}

\def\reff#1{{\rm(\ref{#1})}}

\def\beqs{\begin{eqnarray*}}
\def\enqs{\end{eqnarray*}}
\def\beq{\begin{eqnarray}}
\def\enq{\end{eqnarray}}

\date{}

\begin{document}

\date{\today}

\title{Viscosity characterization of the value function of an investment-consumption problem in presence of illiquid assets}
\author{Salvatore Federico\footnote{Dipartimento di Economia, Management e Metodi Quantitativi, Universit\`a di Milano, Italy. E-mail: \texttt{salvatore.federico@unimi.it} (Part of this research was done when this author was post-doc at  the LPMA - Universit\'e Paris 7 Diderot and member of Alma Research (Paris).)}  \and Paul Gassiat\footnote{Institut f\"ur Mathematik, TU Berlin. E-mail: \texttt{gassiat@math.tu-berlin.de}}}
    \maketitle


\begin{abstract}
\noindent We study a problem of optimal investment/consumption over an infinite horizon in a market consisting of a liquid and an illiquid asset. The liquid asset is observed and can be traded continuously, while the illiquid one can only be traded and observed at discrete random times corresponding to the jumps of a Poisson process.
The problem is a nonstandard mixed discrete/continuous optimal control problem which we face by the dynamic programming approach. The main aim of the paper 
is to prove that the value function is the unique viscosity solution of an associated HJB equation. We then use such result to build a numerical algorithm allowing to approximate the value function and so to    measure the cost of illiquidity.
\end{abstract}


\bigskip

\section{Introduction}
We study a problem of optimal investment/consumption over an infinite horizon in a market consisting of a liquid and an illiquid asset. The liquid asset is observed and can be traded continuously, while the illiquid one can only be traded and observed at discrete random times corresponding to the jumps of a Poisson process with intensity $\lambda>0$. 

This kind of model for the illiquid asset has been already proposed in the literature, e.g. in the  papers  \cite{mat06, phatan08, RZ02}, which deal just with an illiquid asset. Within this setting, the paper \cite{APW} introduces in the market model  also a liquid asset correlated with the illiquid one.\footnote{{Also in  \cite{ST04} the market is composed by a liquid and an illiquid asset. However, there the problem is over a finite horizon and the illiquid asset cannot be traded: the wealth held in the illiquid asset enters just in the optimization functional at the terminal date.}}
  However, in the latter paper it is assumed  full information on the state of the  illiquid asset, differently from \cite{phatan08} where the illiquid asset can be observed only at the trading dates.
Here we consider the point of view of \cite{phatan08}, which seems more realistic, and assume that the illiquid asset can be only observed at the trading dates. Another main difference with respect to  \cite{APW} is that here we consider general utility, so we cannot perform a reduction of variable by homogeneity, which in \cite{APW} is allowed by the choice of CRRA utility.

In Section 2,  we set the  problem as a mixed discrete/continuous stochastic optimal control problem. Such a problem is not standard in the theory of optimal stochastic control. Thus, in Section 3 - following the approach of \cite{phatan08} - by means of a specific dynamic programming principle we reduce  the  control problem between trading times to a  continuous time-inhomogeneous problem. Then we state the main result  of the paper (Theorem \ref{thmVisc}), providing the characterization of  the value function of the reduced problem as  unique  continuous viscosity solution of the associated Hamilton-Jacobi-Bellman (HJB)  equation. 
This result allows us to perform in Section 4 a numerical iterative scheme to approximate the value function, which is not straighforward due to the presence of a nonlocal term in the HJB equation.
In Section 5 we exploit the results obtained  providing some first answers to the problem: we describe the structure of the optimal allocation policy in the illiquid asset and give a numerical measure of the cost of illiquidity. 

{In order to go further into the solution and state the optimal allocation in the liquid asset as well as the optimal consumption rate, one has to prove regularity results for the value function. These results are the object, in the case of power utility, of the companion paper \cite{FGGfin}, where more numerical tests are performed.}


\section{Market model and optimization problem}
Let us consider a complete filtered probability space $(\Omega,\Fc,(\Fc_t)_{t\geq 0},\P)$ satisfying the usual conditions, on which there are defined:
\begin{itemize}
\item[-]A Poisson process $(N_t)_{t\geq 0}$, with intensity $\lambda>0$; we denote by $(\mathcal{N}_t)_{t\geq 0}$ the filtration generated by this process and  by
$(\tau_n)_{n\geq 1}$ its  jump times; moreover we set $\tau_0=0$.
\item[-]Two independent standard Brownian motions $(B_t)_{t\geq 0}$ and $(W_t)_{t_\geq 0}$, independent also of the Poisson process $(N_t)_{t\geq 0}$; we denote by $(\mathcal{B}_t)_{t\geq 0}$  and $(\mathcal{W}_t)_{t\geq 0}$ the filtration generated by $B$ and $W$ respectively.
\end{itemize}
The market model we consider on this probability space is composed by a riskless asset with constant return rate, which for sake of simplicity we consider equal to $0$, and 
two risky assets with correlation $\rho\in(-1,1)$: \begin{itemize}
\item[-]
A liquid risky asset that can be traded continuously; it is described by a stochastic process denoted by $L_t$ whose dynamics is
\beqs
dL_t& = &b_L L_tdt + \sigma_L L_tdW_t,
\enqs
where $b_L\in\R$ and $\sigma_L>0$.
\item[-]An illiquid risky asset that can only be traded at the trading times $\tau_n$; it is described by a stochastic process denoted  by $I_t$, whose dynamics is
\beqs
dI_t& =& b_I I_t dt + \sigma_I I_t\left(\rho dW_t + \sqrt{1 - \rho^2} dB_t\right),
\enqs
where $b_I\in\R$ and $\sigma_I>0$.
\end{itemize}
Without loss of generality we assume $L_0 = I_0=1$. 
Define the $\sigma$-algebra
$$\mathcal{I}_t\ \ = \ \ \sigma\left(I_{\tau_n}\mathbf{1}_{\{\tau_n\leq t\}}, \ n\geq 0\right), \ \ \ t\geq 0.$$
Moreover define the  filtration
$$\mathbb{G}^0\ = \ (\Gc^0_t)_{t\geq 0}; \ \ \ \ \ \ \ \ \ \mathcal{G}^0_t\ = \ \mathcal{N}_t\vee \mathcal{I}_t\vee \mathcal{W}_t \  = \  \sigma(\tau_n,I_{\tau_n}; \tau_n\leq t) \vee \mathcal{W}_t.$$
The observation filtration we consider is
$$\mathbb{G}\  =\  (\mathcal{G}_t)_{t\geq 0}; \ \  \ \ \ \ \mathcal{G}_t\  = \ \mathcal{G}^0_t\vee \sigma(\P\mbox{-null sets}).$$
This means that at time $t$ the agent knows the past of the liquid asset up to time $t$, the trading dates of the illiquid assets occurred before $t$, and the values of the illiquid asset at such trading dates.\\\\
In the setting above, we define a set of admissible trading/consumption strategies in the following way.
Consider all the triplets of processes $(c_t,\pi_t,\alpha_k)$ such that:
\begin{itemize}
	\item[(h1)] $c=(c_t)_{t\geq 0}$ is a continuous-time nonnegative process $(\Gc_t)_{t\geq 0}$-predictable and with  locally integrable trajectories; $c_t$ represents the \emph{consumption rate} at time $t$.
	\item[(h2)]$\pi = (\pi_t)_{t\geq 0}$ is a continuous-time process $(\Gc_t)_{t\geq 0}$-predictable with  locally square integrable trajectories; $\pi_t$ represents the \emph{amount of money invested in the liquid asset} at time $t$.
	\item[(h3)] $\alpha=(\alpha_k)_{k\in\mathbb{N}}$, is a discrete-time process, where $\alpha_k$ is  $\Gc_{\tau_k}$-measurable; $\alpha_k$ represents the \emph{amount of money invested in the illiquid asset} in the interval  $(\tau_k,\tau_{k+1}]$.
\end{itemize}
Given a triplet  $(c,\pi,\alpha)$ satisfying the requirements (h1)--(h3) above and an initial wealth $r\geq 0$, we can consider the process $R_t$ representing the wealth associated to such strategy. Its dynamics can be defined by recursion on $k\geq 0$ by
\beqs
R_0 &=& r, \\
R_t &=& R_{\tau_k} + \int_{\tau_k}^{t} ( -c_s ds + \pi_s (b_L ds + \sigma_L dW_s))  + \alpha_k \left(\frac{I_{t}}{I_{\tau_k}} -1\right), \ \ t\in(\tau_k,\tau_{k+1}].
\enqs
As a class of admissible controls we consider all the triplets  of processes $(c,\pi,\alpha)$ satisfying the measurability and integrability conditions above and such that the corresponding wealth process $R_t$ is nonnegative (no-bankruptcy constraint). 
The class of admissible controls depends on the initial wealth $R_0=r$. We denote this class by $\mathcal{A}(r)$, noticing that it is not empty for every $r\geq 0$, as $(0,0,0)\in\mathcal{A}(r)$.
\\\\
Given a utility function $U:\R_+ \longrightarrow\R$, the optimization problem we want to solve  is
\beq\label{optpr}
\mbox{Maximize} \ \ \ \E \left[ \int_0^{\infty} e^{-\beta s} U(c_s) ds \right], \ \ \ \mbox{over} \  {(c,\pi,\alpha) \in \Ac(r)}.
\enq
\begin{assumption}\label{ass:U}
The  utility function $U$ is continuous, nondecreasing, concave and bounded from below (without loss of generality we assume that  $U(0)=0$). Moreover, it satisfies the growth condition, for some $K_U>0$,
\beq\label{growthU}
U(c)&\leq& K_U\,c^p.
\enq
\end{assumption}
\begin{assumption}\label{ass:beta}
We assume that
\beq \label{ineqBeta}
\beta\; > \; k_{p}, 
\enq
where
\beqs 
k_{p}&:=& \sup_{u_L\in \R ,u_I \in [0,1]} \left\{p (u_L b_L + u_I b_I) - \frac{p(1-p)}{2} ( u_L^2 \sigma_L^2 + u_I^2 \sigma_I^2 + 2 \rho u_L u_I \sigma_L \sigma_I)\right\}. 
\enqs
\end{assumption}
\medskip
For convenience we set
\beqs
\tilde{k}_{p} &:=& \sup_{u_I \in [0,1]} \left\{p (b_I - \frac{\rho b_L \sigma_I}{\sigma_L}) u_I - \frac{p(1-p)}{2} \sigma_I^2 (1-\rho^2) u_I^2\right\}.
\enqs
so that
\beqs
k_p&=& \frac{p}{2(1-p)} \frac{b_L^2}{\sigma_L^2} + \tilde{k}_{p}.
\enqs

\begin{remark}\label{remMerton}
The assumption on $\beta$ is related to the investment/consumption problem with the same assets but in a liquid market.
Let $p\in(0,1)$ and consider an agent with initial wealth $r$, consuming at rate $c_t$ and investing in $L_t$ and $I_t$ continuously with respective proportions $u^L_t$ and $u^I_t$ and under the constraint that $u^I_t \in [0,1]$. Suppose, moreover, that the preferences of the agent are represented by the utility function  $U^{(p)}(c)=c^p/p$, with $p\in (0,1)$. Let us  denote by $\Ac_{{Mert}}(r)$ the set of strategies keeping the wealth nonnegative and define the value function
\beq
V^{(p)}_{Mert}(r) &=& \sup_{(u^L,u^I,c) \in \Ac_{Mert}(r)} \E \left[\int_0^\infty e^{-\beta t} U^{(p)}(c_t) dt\right], \label{defVM}
\enq

 This is a constrained Merton problem which dominates our problem, in the sense that $V^{(p)}_{Mert}(r)$ is higher of  the optimal value of our problem, up to the multiplicative constant $K_U$ of \eqref{growthU}.  One can see (for instance by solving the HJB equation) that $V^{(p)}_{Mert}$ is finite if and only if \reff{ineqBeta} is satisfied
 and that in this case
\beq
V^{(p)}_{Mert}(r) &=&  \left(\frac{1-p}{\beta - k_{p}}\right)^{1-p} r^p. \label{ExplVM}
\enq
Therefore, condition \eqref{ineqBeta} guarantees together with \eqref{growthU} finiteness for our problem too. 
\end{remark}

\section{Dynamic programming and HJB equation}
Let us denote by $V$ the value function of  the stochastic control problem \eqref{optpr}:
\beqs
V(r) &=& \sup_{(c,\pi,\alpha) \in \Ac(r)} \E \left[ \int_0^{\infty} e^{-\beta s} U(c_s) ds \right].
\enqs
\begin{proposition}\label{prop:V}
$V$ is everywhere finite, concave, $p$-H\"older continuous  and nondecreasing.  Moreover
\beq\label{estimV}
{V}(r)&\leq & K_V r^p, \ \ \ \mbox{for some} \ K_V>0.
\enq
\end{proposition}
{\bf Proof.} As we have already observed in Remark \ref{remMerton}, finiteness and \eqref{estimV} follow from \eqref{growthU} and \eqref{ineqBeta}, by comparing with a constrained Merton problem.

Concavity of $V$ comes from concavity of $U$ and linearity of the state equation by standards arguments. Also monotonicity is consequence of standard arguments due to  monotonicity of $U$. Finally, $p$-H\"older continuity follows from concavity and monotonicity of $V$, and from \eqref{estimV}. \hfill$\square$\\\\
Following \cite{phatan08}, we state a suitable  Dynamic Programming Principle (DPP) to reduce our mixed discrete/continuous problem to a standard one between two trading times.
\begin{proposition}[DPP] \label{propDPP}
We have the following equality:
\beq \label{eqDPP0}
V(r) &=& \sup_{(c,\pi,\alpha)\in \Ac(r)} \E\left[ \int_0^{\tau_1} e^{-\beta s} U(c_s) ds + e^{-\beta \tau_1} V\left(R_{\tau_1} \right) \right].
\enq
\end{proposition}

{\bf Proof.} The proof is long and technical, but similar to the one in \cite{phatan09} and we omit it for brevity. Note however that, unlike in \cite{phatan09}, there is some additional information between $0$ and $\tau_1$ brought by $W$, so the ``shifting" procedure is slightly more technical to achieve. One can see, for instance, Appendix B in \cite{gasgozpha11} for details on the shifting procedure when there is a random process bringing an extra information between $0$ and $\tau_1$.
\ep
\\\\
We can use this DPP to relate our original problem into a standard continuous-time control problem.
First of all, letting $\mathcal{M}(\R_+;\R)$ denote the space of measurable functions from $\R_+$ to $\R$, we define
 the linear operator 
\begin{equation}\begin{array}{cccc} \label{defG}
G: &  \mathcal{M}(\R_+;\R)&\longrightarrow &\mathcal{M}([0,+\infty)\times \R_+^2;\R)\\
 &\psi&\longmapsto &  G[\psi](t,x,y):= \E\left[\psi(x+yJ_t) \right],
\end{array}
\end{equation}
where
\beq\label{EJ}
\frac{dJ_t}{J_t}& =&   \left (b_I - \rho b_L \frac{\sigma_I}{\sigma_L}\right) dt + \sigma_I \sqrt{1-\rho^2} dB_t, \ \ \ \ J_0\ =\ 1.
\enq
For each $x\geq 0$ and $t\geq 0$, let $\Ac_{t}(x)$ be the set of couples of stochastic processes $(c_s,\pi_s)_{s\geq t}$  such that
\begin{itemize}
\item[-]$(c_s)_{s\geq t}$ is $(\mathcal{W}_s)_{s\geq t}$-predictable,  nonnegative  and has locally integrable trajectories;
\item[-]$(\pi_s)_{s\geq t}$ is $(\mathcal{W}_s)_{s\geq t}$-predictable   and has locally square integrable trajectories;
\item[-]
 $x + \int_t^{T} (-c_s ds +\pi_s (b_L ds + \sigma_L dW_s)) \geq 0,$  for all $T\geq t.$
 \end{itemize}
Using \eqref{eqDPP0}, one can show (see \cite{FGGfin}) that
\beq\label{vhatv}
V(r) &=& \sup_{0\leq a\leq r} \widehat V(0,r-a,a),\ \ \ \ \ r\geq 0,
\enq
where 
\beq\label{hatv}
\widehat{V}(t,x,y)&=& \sup_{ ({c},{\pi}) \in \Ac_t(x)}
\mathcal{J}(t,x,y;{c},{\pi}), \ \ \ \ (t,x,y)\in \R_+^3, \label{defvhat}
\enq
and
\beq\label{JV}
\mathcal{J}(t,x,y;{c},{\pi})& =& \E\int_t^\infty e^{- (\beta+\lambda) (s-t)} \left( U({c}_s) + \lambda G[V]\left(s, {X}_s^{t,x,{\pi},{c}}, {Y}_s^{t,y}\right) \right)ds,
\enq
with  $(X_s^{t,x,{c},{\pi}})_{s\geq t}, \ ({Y}_s^{t,y})_{s\geq t}$  solutions  to the SDEs
\begin{eqnarray}\label{XYt}
d{X}_s &=& -{c}_s ds + {\pi}_s (b_L ds + \sigma_L dW_s), \ \ \ \ \ \ X^{t,x,c,\pi}_t\ =\ x,\\
\label{XY2t}
d{Y}_s &=&\rho{Y}_s\left( \frac{b_L \sigma_I}{\sigma_L} dt+  \sigma_I dW_s\right), \ \ \ \  \ \ Y^{t,y}_t\ = \ y.
\end{eqnarray}
We notice that the problem of optimizing the functional above is not autonomous due to the dependence of $G[V]$ on time. 

Associating to every locally bounded function $\hat{v}: \R_+^3\rightarrow\R$ the function $\Hc \hat{v}:\R\rightarrow\R$ defined  by 
\beqs
[\Hc \hat{v}](r) &=& \sup_{0 \leq a \leq r} \hat{v}(0,r-a,a),
\enqs
by  the arguments above we may rewrite the original problem as
\beq\label{vhatv}
V (r)&=& [\Hc \widehat V](r).
\enq
The problems \eqref{hatv} and \eqref{vhatv} are coupled in the sense that $\widehat{V}$ is expressed in terms of $V$ in \eqref{hatv} and, viceversa, $V$ can be expressed in terms of $\widehat{V}$ by \eqref{vhatv}. 
\subsection{Properties of $\widehat{V}$}
In this subsection we prove some qualitative properties of the value function $\widehat V$. First, we start by studying some properties of the operator  $G$.
\begin{proposition}\label{prop:G}
\begin{itemize}
\item[]
\item[(i)] $G$ is  well defined on the set of measurable functions with polynomial growth.
\item[(ii)] $G$ is  positive, in the sense that it maps positive functions into positive ones. 
\item[(iii)] $G$ maps increasing functions to functions which are increasing with respect to both $x$ and $y$.
\item[(iv)] $G$ maps concave   functions to functions which are concave with respect to   $(x,y)$.
\item[(v)]  If $\psi(r)=r^p,$ $p\in (0,1)$, then
\begin{equation}\label{eqGgrowthG}
0\ \leq\  G[\psi](t,x,y)\ \leq \ e^{\tilde{k}_{p} t}(x+y)^p, \ \ \ \forall t\geq 0,\ \forall (x,y)\in\R^2_+.
\end{equation}
\item[(vi)]Let $p \in(0,1]$ and $\psi$ a $p$-H\"older continuous function. Then there exists some constant $C \geq 0$ such that  for all $t\geq 0$, $x,x',y,y'\geq 0$, and $0< h \leq 1$, 
\beq
|G[\psi](t,x,y) - G[\psi](t,x',y)| &\leq& C |x-x'|^p, \label{HoldGx} \\
|G[\psi](t,x,y) - G[\psi](t,x,y')|&\leq& C e^{\tilde{k}_{p} t} |y-y'|^p, \label{HoldGy}  \\
|G[\psi](t,x,y) - G[\psi](t+h,x,y)| &\leq& C e^{\tilde{k}_{p} t} y^p h^{p/2}, \label{HoldGt}
\enq
\end{itemize}
\end{proposition}
{\bf Proof.} (i)--(iv)  are straightforward.

(v). 
If $x=y=0$ the claim is obvious, so we assume $x+y>0$. 
By a straightforward application of It\^o's formula and the definition of $\tilde{k}_{p}$, we see that $(e^{-\tilde{k}_{p}t}(x+yJ_t)^p)_{t\geq 0}$ is a supermartingale, which implies \reff{eqGgrowthG}.

 (vi). \reff{HoldGx} is obvious, and \reff{HoldGy} follows directly from (v). To prove \reff{HoldGt}, fix $(t,x,y)\in\R^3_+$ and $h\in(0,1]$. We can write for some $C>0$
\beq\label{mmmz}
|G[\psi](t,x,y) - G[\psi](t+h,x,y)| &\leq& C y^p \E \left[ |J_t-J_{t+h}|^p\right] \nonumber\\
&=& C y^p \E \left[ |J_t|^p\right] \E\left[ |1-J_{h}|^p\right] \\
&\leq& C e^{\tilde{k}_{p}t} y^p \E\left[ |1-J_{h}|^p\right]\nonumber.
\enq
Now we have $J_{h} = e^{\alpha h + \beta \sqrt{h} N}$, where $\alpha$, $\beta$ are constants and where $N \sim \Nc(0,1)$. Since $|e^\xi - 1| \leq |\xi| (e^\xi + 1)$ for all $\xi$ $\in$ $\R$, we obtain for some $C_1>0$
\beq\label{op}
\frac{\E\left[ |1-J_{h}|^p\right]}{h^{p/2}} &\leq& \E \left[ (\alpha \sqrt{h} + \beta N)^p ( e^{\alpha h + \beta \sqrt{h} N} +1)^p \right]
\ \ \leq\ \ C_1 , \;\;\; \mbox{ for } 0< h \leq 1.
\enq
The claim follows combining \eqref{op} with \eqref{mmmz}.
\ep
\begin{lemma} \label{lemGrowthXY}
For $(t,x,y)$ $\in$ $\R^3_+$, $(c,\pi)$ $\in$ $\Ac_t(x)$, $p$ $\in$ $(0,1)$,
\beq \label{ineqGrowthXY}
\E \left[({X}_s^{t,x,{c},{\pi}}+{Y}_s^{t,y})^p\right] &\leq& e^{ \frac{p}{1-p}\frac{b_L^2}{2 \sigma_L^2} (s-t)} (x+y)^p,\ \ \ \ \forall s\geq t.
\enq

\end{lemma}
{\bf Proof.}
Fix $(t,x,y)\in\R_+^3$ and $(c,\pi)$ $\in$ $\Ac_t(x)$.
First of all we notice that by standard comparison of SDE's we have
\beq\label{X0Xc}
X^{t,x,c,\pi}&\leq& X^{t,x,0,\pi}.
\enq
On the other hand we have
\beqs
\frac{d({X}_s^{t,x,0,{\pi}}+{Y}_s^{t,y})}{{X}_s^{t,x,0,{\pi}}+{Y}_s^{t,y}} &=& U_s (b_L ds + \sigma_L dW_s),
\enqs
where
\beqs
U_s&=& \frac{\pi_s  + \rho \frac{ \sigma_I}{\sigma_L} Y_s^{t,y}}{X_s^{t,x,0,{\pi}} + Y_s^{t,y}}.
\enqs
Noticing that $\frac{p}{1-p}\frac{b_L^2}{2 \sigma_L^2} = \sup_{u \in \R} \left\{p b_L u - \frac{p(1-p)}{2} \sigma_L^2 u^2\right\}$,
it is then a straightforward application of It\^o's formula to check that $\left(\exp(- \frac{p}{1-p}\frac{b_L^2}{2 \sigma_L^2}(s-t)) ({X}_s^{t,x,0,{\pi}}+{Y}_s^{t,y})^p\right)_{s\geq t}$ is a local supermartingale, and, being nonnegative, a true supermartingale. Therefore, we have the claim for $c= 0$. The general claim follows from \eqref{X0Xc}.
\ep
\begin{proposition}\label{prop:Vhat}
 $\widehat{V}(t,\cdot)$ is concave with repect to $(x,y)$ and  nondecreasing with respect to  $x$ and $y$  for every $t\geq 0$. Moreover it satisfies the boundary condition
 \beq \label{Bndryv}
\widehat{V}(t,0,y)&=& \E \int_t^\infty e^{-(\beta+\lambda)(s-t)} \lambda G[V](s,0,{Y}^{t,y}_s) ds, \ \ \ \forall t\geq 0, \ \forall y\geq 0.
\enq
In particular, since by Assumption \ref{ass:U} it is $U(0)=0$, due to Proposition \ref{prop:G}(v) we have
\beq \label{Bndry2}
\widehat{V}(t,0,0)&=&0, \ \ \ \forall t\geq 0.
\enq
Finally, $\widehat V$ is continuous on $\R_+^3$, and satisfies for some $K_{\widehat{V}}>0$ the growth condition
\beq \label{Growthvhat}
0\ \ \leq\ \  \widehat V(t,x,y) &\leq& K_{\widehat V} e^{\tilde{k}_{p} t} (x+y)^p, \ \ \ \ \forall  (t,x,y) \in \R^3_+.
\enq
\end{proposition}

\textbf{Proof.} 
\emph{Concavity and monotonicity.}
Since $V$ is concave and nondecreasing, by Proposition \ref{prop:G}(iii,\,iv), $G[V](t,\cdot)$ is concave in $(x,y)$ and nondecreasing in $x,y$ on $\R_+^2$.  Then concavity and monotonicity of $\widehat V$ follow by standard arguments, considering also the linearity of the SDE's \eqref{XYt}-\eqref{XY2t}.

\ni
\emph{Boundary condition.} Equality \eqref{Bndryv} is due to the fact that $\mathcal{A}_t(0)=\{(0,0)\}$, so
\beqs
\widehat{V}(t,0,y)&=&\mathcal{J}(t,0,y;0,0)\ \ = \ \ \E \int_t^\infty e^{-(\beta+\lambda)(s-t)} \lambda G[V](s,0,{Y}^{t,y}_s) ds.
\enqs
\smallskip
\noindent
\emph{Continuity.} We prove the continuity of $\widehat V$ in several steps.

1)  Continuity of $\widehat{V}(t,\cdot)$ in $(0,+\infty)^2$ follows from concavity.

2) Here  we prove the continuity of $\widehat{V}(t,\cdot,y)$ at $x=0^+$. First of all notice that \eqref{Bndryv} holds at $x=0$, so using  monotonicity of $V$  and \ref{prop:G}-(iii) we get
 \beq\label{xc}
0&\leq& \mathcal{J}(t,x,y;0,0)- \mathcal{J}(t,x,0;0,0)\ \ \leq\ \  \widehat{V}(t,x,y)- \widehat{V}(t,x,0).
\enq
On the other hand,
using   H\"older continuity of $V$ and  \eqref{HoldGx},  we have for some $K>0$ and all $({c},{\pi})\in\mathcal{A}_t(x)$
\begin{multline*}
 \mathcal{J}(t,x,y;{c},{\pi})-V(t,0,y)\\\leq  \ \E\left[ \int_t^\infty e^{-(\beta+\lambda)(s-t)} \left\{ U({c}_s) \ + \lambda\,  \big|G[V](s,{X}^{t,x,{c},{\pi}}_s,{Y}^{t,y}_s) - G[V](s,0,{Y}^{t,y}_s)\big| \right\} ds \right]  \\
\leq  \ \E\left[ \int_t^\infty e^{-(\beta+\lambda)(s-t)} \left( U({c}_s) +    K|{X}^{t,x,{c},{\pi}}_s|^p\right) ds\right].
\end{multline*}
Taking the supremum over  $({c},{\pi})\in\mathcal{A}_t(x)$  and combining with \eqref{xc} we get
\beq\label{ps}
0 \,\leq \, \widehat{V}(t,x,y) - \widehat{V}(t,0,y) \,\leq \, \sup_{({c},{\pi}) \in \Ac_t(x)}  \E\left[ \int_t^\infty e^{-(\beta+\lambda)(s-t)} \left( U({c}_s) +    K|{X}^{t,x,{c},{\pi}}_s|^p\right) ds\right].
\enq
We have to estimate the right handside of \eqref{ps}.
By definition of $\Ac_t(x)$, we have
\beq\label{sa1}
0 &\leq &{X}^{t,x,{c},{\pi}}_s \ \ =\ \ x + \int_t^s {\pi}_u \frac{dL_u}{L_u} - \int_t^s {c}_u du.
\enq
Denoting by $\Q^L$ the probability with density process given by $Z_t= \exp\left(- \frac{b_L^2}{2 \sigma_L^2} t - \frac{b_L}{\sigma_L} W_t\right)$, $L$ is a $\Q^L$-martingale. The process  ${X}^{t,x,{c},{\pi}}$ is then a $\Q^L$-local supermartingale and, being bounded from below, it is a true $\Q^L$-supermartingale. Hence, we have $\E[Z_s {X}^{t,x,{c},{\pi}}_s ] \leq x$. Now, writing $|{X}^{t,x,{c},{\pi}}_s|^p = |Z_s {X}^{t,x,{c},{\pi}}_s|^p Z_s^{-p}$, by H\"older's inequality we get
\beq
\E[| {X}^{t,x,{c},{\pi}}_s|^p] &\leq& \E[Z_s   {X}^{t,x,{c},{\pi}}_s]^p \,\,\E[Z_s^{- \frac{p}{1-p}}]^{1-p}
\ \ \leq \ \ x^p \exp\left( \left( \frac{p}{1-p} \frac{b_L^2}{2 \sigma_L^2}\right) s\right). \label{ineqx2}
\enq
Note also that, since $\int_t^\infty e^{-(\beta+\lambda)(s-t)} U({c}_s) ds$ is the utility obtained by the agent trading only in $L$, we have by \reff{estimV}
\beq
\sup_{({c},{\pi}) \in \Ac_t(x)}  \E\left[ \int_t^\infty e^{-(\beta+\lambda)(s-t)} U({c}_s) ds \right] &\leq& V(x) \ \
\leq\ \ K_V x^p, \label{ineqx3}
\enq
Combining \eqref{ps}, \eqref{ineqx2}, \eqref{ineqx3}, and using \eqref{ineqBeta}, we get for some
 $K>0$
\beq\label{ps1}
0&\leq &\widehat{V}(t,x,y) - \widehat{V}(t,0,y)\ \ \leq \ \ K x^p,
\enq
and we conclude.

3) Here we prove the continuity of $\widehat{V}(t,x,\cdot)$ at $y=0^+$.
Using  monotonicity of $V$  and Proposition \ref{prop:G}(iii) we get
\beq\label{estbe}
0& \leq& \mathcal{J}(t,x,y;{c},{\pi})- \mathcal{J}(t,x,0;{c},{\pi})
\enq
On the other hand, using H\"older continuity of $V$, \eqref{HoldGy}   and \reff{ineqGrowthXY},  we have for some $K>0$ and for all $({c},{\pi})\in\mathcal{A}_t(x)$
\beq\label{estab}
 \mathcal{J}(t,x,y;{c},{\pi})- \mathcal{J}(t,x,0;{c},{\pi})&\leq& K  e^{\tilde{k}_{p}t}\int_t^\infty e^{-(\beta+\lambda-\tilde{k}_{p})(s-t)} \lambda \E[(Y^{t,y}_s)^p] ds\nonumber \\
&\leq &K  e^{\tilde{k}_{p}t} y^p \int_t^\infty e^{-(\beta+\lambda- k_{p})(s-t)}  \lambda ds\\
&
=& K \frac{\lambda}{\beta+\lambda-k_{p}} e^{\tilde{k}_{p}t} y^p.\nonumber
\enq
 Therefore, taking the supremum over  $({c},{\pi})\in\mathcal{A}_t(x)$ in \eqref{estab} and combining with \eqref{estbe}, we get
\beq\label{esthatV0}
0&\leq &\widehat{V}(t,x,y)- \widehat{V}(t,x,0)\ \ \leq\ \ \frac{K \lambda}{\beta+\lambda- k_{p}} e^{\tilde{k}_{p}t} y^p,
\enq
and we conclude.

4) Since \eqref{esthatV0} and \eqref{ps1} are uniform estimates in $x,y$ respectively, combining with the continuity on the lines provided by  items 2) and 3), we get the joint continuity of $\widehat{V}$ with repect to $(x,y)$ at the boundary $\{(x,y)\in\mathbb{R}^2_+ \ | \ x=0 \ \mbox{or} \ y=0\}.$

5) Here we prove $p/2$-H\"older continuity of $\widehat{V}(\cdot,x,y)$.
Let $t,t'\geq 0$ and suppose that $t'=t+h$ for some $0< h\leq 1$. One can associate to each  $({c}^t_s,{\pi}^t_s)_{s\geq t}\in \mathcal{A}_t(x)$ a control $({c}^{t'}_s,{\pi}^{t'}_s)_{s\geq t'}\in \mathcal{A}_{t'}(x)$  with the same law and viceversa (see \cite[Th.\,2.10, Ch.\,1]{YZ}).
Given that and considering  \eqref{HoldGt} and \reff{ineqGrowthXY}, we have for some $K>0$
\beqs
&& |\mathcal{J}(t,x,y,{c}^t,{\pi}^t) -\mathcal{J} (t',x,y,{c}^{t'},{\pi}^{t'})| \\
 &\leq& \E \int_t^\infty e^{-(\beta+\lambda)(s-t)} \lambda\  \left|G[V](s,{X}^{t,x,{c}^t,{\pi}^t},{Y}^{t,y}_s)ds -
G[V](s+h,{X}^{t,x,{c}^{t},{\pi}^{t}},{Y}^{t,y}_s)\right|ds \\
&\leq& K \E  \int_t^\infty e^{-(\beta+\lambda)(s-t)}\lambda h^{p/2} e^{\tilde{k}_{p}s} |Y^{t,y}_s|^p ds \\
&\leq& K \frac{\lambda}{\beta + \lambda - k_{p}}\,e^{\tilde{k}_{p}t} h^{p/2}.
\enqs
Passing to the supremum over $({c}^t_s,{\pi}^t_s)_{s\geq t}\in \mathcal{A}_t(x)$ (respectively over $({c}^{t'}_s,{\pi}^{t'}_s)_{s\geq t'}\in \mathcal{A}_t'(x)$), we get for some $K>0$
\beq\label{xm}
|\widehat V(t,x,y) - \widehat V(t+h,x,y)| &\leq& K e^{\tilde{k}_{p}t} y^p h^{p/2}.
\enq
Hence $\widehat{V}$ is locally $p/2$-H\"older with respect to $t$.

6) Putting together all the information collected we get continuity of $V$ on $\R_+^3$.

\smallskip
\noindent
\emph{Growth condition}. Condition \reff{Growthvhat} is proved by combining \reff{ps1}, \reff{esthatV0} and \reff{Bndry2}.
\ep
\subsection{HJB equation: viscosity characterization of $\widehat{V}$}
By standard arguments of stochastic control (see e.g. \cite[Ch.\,4]{YZ}), we can associate to $\widehat{V}$ an HJB equation, which in this case reads as
\beq\label{HJBv}
- \hat{v}_t  + (\beta+\lambda) \hat{v}  - \lambda G[\Hc \hat{v}]- \sup_{c\geq 0, \pi \in \R} H_{cv}(y,D_{(x,y)} \hat{v},D^2_{(x,y)} \hat{v};\ c,\pi)&
 =& 0,
\enq
where for $(y,q,Q)$ $\in$ $\R_+ \times \R^2 \times \Sc_2$ (where $\Sc_2$ denotes the space of symmetric $2 \times 2$ matrices), $c\geq 0, \ \pi \in \R$, the function $H_{cv}$ is defined by
\begin{multline*}
H_{cv}(y,q,Q;c,\pi)\\
 =  \ \ U(c) + (\pi b_L - c) q_1 +  \frac{\rho b_L \sigma_I}{\sigma_L}  y q_2+ \frac{\sigma_L^2 \pi^2}{2} Q_{11} +   \pi \rho \sigma_I \sigma_L y Q_{12}+  \frac{\rho^2 \sigma_I^2}{2} y^2 Q_{22}.
\end{multline*}
Note that $\sup_{c\geq 0, \pi \in \R} H_{cv}(y,q,Q;c,\pi)$ is finite if $q_1 > 0$, $Q_{11}<0$, in which case we have
\begin{multline*}
\sup_{c\geq 0, \ \pi \in \R} H_{cv}(y,q,Q;c,\pi)\; = \;\widetilde{U}(q_1) - \frac{(b_L q_1 + \rho \sigma_L \sigma_I y Q_{12})^2}{2 \sigma_L^2 Q_{11}}+  \frac{\rho b_L \sigma_I}{\sigma_L}  y q_2 +  \frac{\rho^2 \sigma_I^2}{2} y^2 Q_{22}.
\end{multline*}
Let us denote by $X = (x,y)$ vectors in $\R_+^2$. We are going to prove that $\widehat V$ is the unique constrained viscosity solution to \eqref{HJBv} according to the following definition.

\begin{definition}\label{Def:visc}
\rm
\ni \textbf{(1)}  Given $\hat v$ a continuous function on $\R_+^3$,  the parabolic {\it superjet} of $\hat v$  at
$(t,X)$ $\in$ $\R_+^3$ is defined by:
\begin{multline*}
\Pc^{1,2,+}\hat v(t,X) =  \Big\{(r,q,Q) \in \R \times \R^2 \times \Sc^2 \mbox{   such that  } \\
\hat v(s,X') \leq \hat v(t,X) + r(s-t) + \left\langle q, X' -X\right\rangle  
                       + \frac{1}{2}\left\langle Q(X'-X), X' -X\right\rangle\ + o\big({|s-t|+}\big|X'-X\big|^2 \big) \Big\},
\end{multline*} 
We define its closure $\overline{\Pc}^{1,2,+}v(t,X)$ as the set  of  elements $(r,q,Q)$ $\in$ $\R \times\R^2\times\Sc^2$ for which there exists a sequence
$(t_m,X_m,r_m,q_m,P_m)_m$ of $\R_+^3\times\Pc^{1,2,+}\hat v(t_m,X_m)$ satisfying  $(t_m,X_m,r_m,q_m,Q_m)$ $\rightarrow$ $(t,X,r,q,Q)$.
We also define the subjets 
$$\Pc^{1,2,-}\hat v(t,X)\ \ = \ \ \Pc^{1,2,+}(-\hat v)(t,X), \ \ \ \  \ \ \ \ \overline{\Pc}^{1,2,-}\hat v(t,X)\ \ =\ \ -\overline{\Pc}^{1,2,+}(-\hat v)(t,X).$$
\smallskip
\ni \textbf{(2)}
We say that a continuous function $\hat v$ is a viscosity subsolution (resp. supersolution) to \eqref{HJBv} at $(t,X)$ $\in$ $\R_+^3$ if
\beqs
- r  + (\beta+\lambda) \hat v(t,X)  - \lambda G[\Hc \hat v](t,X) - \sup_{c\geq 0, \pi \in \R} H_{cv}(y,q,Q;c,\pi) &\leq& 0,
\enqs
for all $(r,q,Q)$ $\in$ $\overline{\Pc}^{1,2,+}\hat v(t,X)$ (resp. $\geq$\,, $\overline{\Pc}^{1,2,-}\hat v(t,X)$).
\smallskip

\ni \textbf{(3)}
We say that a continuous function $v$ is a \emph{constrained viscosity solution} to \reff{HJBv} if it is a subsolution on $\R_+^3$, a supersolution on $[0,+\infty) \times (0,+\infty) \times \R_+$ and satisfies the boundary condition 
\beq\label{bdrcond}
\hat {v}(t,0,y)&=& \E \Big[\int_t^\infty e^{-(\beta+\lambda)(s-t)} \lambda G[\mathcal{H}\hat v](s,0,{Y}^{t,y}_s) ds\Big], \ \ \ \forall t\geq 0, \ \forall y\geq 0.
\enq
\end{definition}
\smallskip
\begin{remark}\label{frontiere}
The concept of constrained viscosity solution we use naturally comes  from the stochastic control problem. The boundaries $\{x=0, \ y\geq 0\}$ and $\{x\geq 0, \ y=0\}$ are both absorbing for the control problem (in the sense that starting from these boundaries, the trajectories of the control problem remain therein), but they have  different features. Indeed starting from the boundary $\{x\geq 0, \ y=0\}$ the control problem degenerates in a one dimensional control problem; the associated HJB equation is nothing else but our HJB equation restricted to this boundary and this is why we require viscosity sub- and supersolution properties to the value function at this boundary. Instead starting from at the boundary $\{x=0, \ y\geq 0\}$ there is no control problem (since $\mathcal{A}_t(0)=\{(0,0)\}$) and the natural condition to impose is a Dirichlet boundary condition.
\end{remark}

\begin{theorem} \label{thmVisc}
$\widehat{V}$ is the unique constrained viscosity solution to  \eqref{HJBv}
satisfying the growth condition \reff{Growthvhat}.
\end{theorem}
{\bf Proof.} The fact that $\widehat{V}$ is a viscosity subsolution on $\R_+^3$ and a viscosity supersolution on $\mathbb{R}_+ \times (0,+\infty)^2$ is standard (see, e.g., \cite[Ch.\,4]{YZ}). The Dirichlet boundary condition \eqref{bdrcond} is verified due to \eqref{vhatv} and \eqref{Bndryv}. The growth condition \reff{Growthvhat} has been already proved in Proposition \ref{prop:Vhat}.

Therefore, it remains to show that $\widehat{V}$ is a supersolution when  $y=0$. In this case, 
the control problem degenerates in a one dimensional one and again standard arguments apply to this control problem, giving the viscosity supersolution property.

Uniqueness is consequence of the comparison principle Proposition \ref{Prop:comparison} below.\hfill$\square$

\begin{proposition}\label{Prop:comparison}
Let $\hat{w}_1$ (resp. $\hat{w}_2$) be a viscosity subsolution (resp. supersolution) to \reff{HJBv} on $\R_+ \times (0,\infty) \times \R_+$. Assume that $\hat{w}_1$, $\hat{w}_2$ satisfy the growth condition \reff{Growthvhat}, and the boundary condition
\beq \label{eqBndrySub}
\hat{w}_1(t,0,y) &\leq& \E \int_t^\infty e^{-(\beta+ \lambda)(s-t)} \lambda G[\Hc \hat{w}_1](s,0,Y^{t,y}_s) ds
\enq
(resp. $\geq$ for $\hat{w}_2$). Then $\hat{w}_1$ $\leq$ $\hat{w}_2$ on $\R_+^3$.
\end{proposition}

{\bf Proof.} \ \textbf{Step 1.} Starting from $\hat{w}_2$, we construct a sequence of supersolutions $(\hat{w}_{2,n})_{n\geq 1}$ that will be used in the next step to show the comparison.
Fix some $p'\in(p,1)$ such that
\beq\label{pas}
\beta &\geq& k_{p'} \;= \;\frac{p'}{1-p'}\frac{b_L^2}{2 \sigma_L^2} + \tilde{k}_{p'}. \label{ineqBetaq}
\enq
Finding such a $p'$ is possible by \reff{ineqBeta} and by the fact that $p'\mapsto k_{p'}$ is continuous. Define
\beqs
f^{p'}(t,x,y) &:=& e^{\tilde{k}_{p'} t} (x+y)^{p'}.
\enqs
We claim that on $\R_+ \times (0,\infty) \times \R_+$
\beq
- f^{p'}_t  + (\beta+\lambda) f^{p'} - \lambda G[\Hc f^{p'}] 
- \sup_{\pi \in \R} H_{cv}(y,D_{(x,y)} f^{p'},D^2_{(x,y)} f^{p'};0,\pi) &\geq& 0. \label{superf}
\enq
Indeed, first we observe that  $G[\Hc f^{p'}]$ $\leq$ $f^{p'}$ by Proposition \ref{prop:G}(v), and then by straightforward computations we can check that
\beqs
\sup_{ \pi \in \R} \left[\pi b_L  f^{p'}_x +  \frac{\rho b_L \sigma_I}{\sigma_L} yf^{p'}_y + \frac{\sigma_L^2 \pi^2}{2} f^{p'}_{xx} +   \pi \rho \sigma_I \sigma_L y f^{p'}_{xy}+ \rho^2 \frac{\sigma_I^2}{2} y^2 \hat{f}^{p'}_{yy}\right]
&=&  \frac{p'}{1-p'} \frac{b_L^2}{2 \sigma_L^2} \,f^{p'}.
\enqs
Hence,  using \reff{ineqBetaq} we obtain \reff{superf}.
Now given an integer $n \geq 1$, consider the function 
$$\hat{w}_{2,n}  \ \  := \ \  \hat{w}_{2} +  \frac{1}{n} f^{p'}.$$
We claim that for any $(t,x,y)\in\R_+\times (0,\infty)\times \R_+$, the function $\hat{w}_{2,n}$ is a supersolution to \reff{HJBv} at $(t,x,y)$.
Indeed, notice that 
$$\Pc^{1,2,-} \hat{w}_{2,n}(t,x,y)\  \ = \ \ \Pc^{1,2,-} \hat{w}_2(t,x,y) + \frac{1}{n}(f_t^{p'}, D_{(x,y)}f^{p'},D^2_{(x,y)}f^{p'})(t,x,y).$$
So, using subadditivity of $\Hc$, linearity of $G$, the fact that  $f^{p'}_x \geq 0$, linearity of $H_{cv}$ in $(q,Q)$ and \reff{superf}, we have for all $(r,q,Q) \in \Pc^{1,2,-} \hat{w}_2(t,x,y)$ 
\beqs
&& - (q + \frac{1}{n}f^{p'}_t(t,x,y))  + (\beta+\lambda)(\hat{w}_2 (t,x,y)+ \frac{1}{n} f^{p'}(t,x,y)) - \lambda G[\Hc(\hat{w}_2 + \frac{1}{n}f^{p'})](t,x,y) \\
&&- \sup_{c \geq 0 ,\pi \in \R} H_{cv}(y,\ q + \frac{1}{n} D_{(x,y)} f^{p'}(t,x,y),\ Q + \frac{1}{n} D^2_{(x,y)} f^{p'}(t,x,y);\,c,\pi) \\
 &\geq& - q   + (\beta+\lambda) \hat{w}_2 (t,x,y) - \lambda G[\Hc \hat{w}_2](t,x,y)- \sup_{c \geq 0 ,\pi \in \R} H_{cv}(y,q,Q;c,\pi)\\
&& +\ \frac{1}{n}\ \Big\{ - f^{p'}_t (t,x,y) + (\beta+\lambda) f^{p'} (t,x,y)- \lambda G[\Hc f^{p'}](t,x,y) \\
&& \ \ \ \ - \sup_{ \pi \in \R} H_{cv}(y,\ D_{(x,y)} f^{p'}(t,x,y),\ D^2_{(x,y)} f^{p'}(t,x,y);\, 0,\pi) \Big\}\ \geq \ 0.
\enqs
This shows that actually  $\hat{w}_{2,n}$ is a supersolution to \reff{HJBv} at $(t,x,y)$ for each $n\geq 1$.
Moreover,
\beqs
\lambda \E \int_t^\infty e^{-(\beta+\lambda)(s-t)} G [\Hc f^{p'} ](s,0,Y^{t,y}_s) ds &\leq& e^{\tilde{k}_{p'} t} y^{p'} \lambda\ \E \int_t^\infty  e^{(-\beta -\lambda + \tilde{k}_{p'})(s-t)} (Y_s^{t,1})^{p'} ds \\
&\leq&  f^{p'}(t,0,y) \lambda \int_t^\infty e^{(-\beta -\lambda + k_{p'})(s-t)} ds \\
&=& \frac{\lambda}{\beta - k_{p'} + \lambda}\ f^{p'}(t,0,y) \\
& \leq &  f^{p'}(t,0,y),
\enqs
where in the second inequality we have used \reff{ineqGrowthXY}.
By subadditivity of $\Hc$ and linearity of $G$, it follows that 
\beq \label{eqBndrySub2}
\hat{w}_{2,n}(t,0,y) &\leq& \E \int_t^\infty e^{-(\beta+ \lambda)(s-t)} \lambda G[\Hc \hat{w}_{2,n}](s,0,Y^{t,y}_s) ds.
\enq
Finally, notice that by the growth condition on $\hat{w}_1$ and $\hat{w}_2$ we have
\beq \label{eqinf}
\lim_{|(t,x,y)| \rightarrow \infty} (\hat{w}_1 - \hat{w}_{2,n})(t,x,y) &=& - \infty.
\enq

\textbf{Step 2.}
We show that for all $n \geq 1$, it is $\hat{w}_1 \leq \hat{w}_{2,n}$ on $\R_+^3$, and thus conclude that $\hat{w}_1 \leq \hat{w}_2$. Fix $n \geq 1$ and define
\beqs
M &:= &\sup_{[0,+\infty) \times \R_+^2} (\hat{w}_1 - \hat{w}_{2,n}).
\enqs
We want to show that $M\leq  0$. By \reff{eqinf} and continuity of $\hat{w}_1, \hat{w}_{2,n}$, we see that, for some $T_0>0$, $\Cc$ a compact subset of $\R_+^2$, and $ (\bar{t},\bar{x},\bar{y})\in [0,T_0]\times \mathcal{C}$,
\beq\label{maxmax}
M &= &\max_{[0,T_0]\times \Cc^2} (\hat{w}_1 - \hat{w}_{2,n})\  \ =\ \ (\hat{w}_1 - \hat{w}_{2,n}) (\bar{t},\bar{x},\bar{y}).
\enq
 We now distinguish between two cases, showing that both of them lead to conclude $M\leq 0$.

\smallskip

\ni \emph{Case 1}: $\bar x = 0$.
First note that $\Hc \hat{w}_1 - \Hc \hat{w}_{2,n}$ $\leq$ $M$. Using the boundary condition \reff{eqBndrySub}, we then have
\beqs
M &=& (\hat{w}_1 - \hat{w}_{2,n}) (\bar{t},0,\bar{y}) \\
&\leq& \E \int_{\bar t}^\infty e^{-( \beta + \lambda )(s-t)}\lambda G[\Hc \hat{w}_1 - \Hc \hat{w}_{2,n}](s,0, Y^{\bar t,\bar y}_s)ds \\
&\leq& \int_{\bar t}^\infty e^{- (\beta + \lambda) (s-t)}\lambda M ds \\
&=& \frac{\lambda}{\beta + \lambda} M,
\enqs
and it follows that $M \leq 0$.

\smallskip
\ni \emph{Case 2}: $\bar x > 0$.
Using viscosity properties of $\hat{w}_1$ and $\hat{w}_{2,n}$, the nonnegativity of an interior maximum may be proved by the  ``doubling of variables" technique as in \cite{CIL92}. 

Define on $[0,T_0] \times \Cc^2$ the function
\beqs
\Phi_\eps(t,X,X') &= &\hat{w}_1(t,X) - \hat{w}_{2,n}(t,X') - \frac{|X-X'|^2}{2 \eps}.
\enqs
Since $\Phi_\eps$ is continuous on the compact set $[0,T_0] \times \Cc^2$, there exists $(t_\eps,X_\eps,X'_\eps)$ such that
\beqs
M_\eps &:=& \sup_{[0,T_0]\times (\Cc)^2} \Phi_\eps = \Phi_\eps(t_\eps,X_\eps,X'_\eps),
\enqs
and a subsequence, still denoted $(t_\eps,X_\eps,X'_\eps)$, converging to some $(\widehat t, \widehat X, \widehat{X}')$. By standard arguments (see e.g. Lemma 3.1 in \cite{CIL92}), we have
\beq
\lim_{\eps \rightarrow 0} \ \frac{|X_\eps-X'_\eps|^2}{2 \eps} &=& 0, \label{limX-X'}
\enq
from which follows that  $\widehat{X}=\widehat{X}'$ and consequently that $(\widehat t,\widehat X)$ is a maximum point of $(\hat{w}_1-\hat{w}_{2,n})$. Hence, without loss of generality  we can take in \eqref{maxmax}
 \beqs
 (\bar{t},\bar{x},\bar{y})&=& (\widehat t,\widehat X).
 \enqs
Now we apply the parabolic Ishii lemma (Th.\,8.3 in \cite{CIL92}) to obtain $r,r'$ $\in$ $\R$, $Q,Q'$ in $\Sc^2$ such that
\beq
\left(r,\frac{X_\eps-X'_\eps}{\eps},Q\right) \ \in \ \bar{\Pc}^{1,2,+}\hat{w}_1(t_\eps,X_\eps), && \!\!\!\!\!\left(r',\frac{X_\eps-X'_\eps}{\eps},Q'\right) \ \in \ \bar{\Pc}^{1,2,-}\hat{w}_{2,n}(t_\eps,X'_\eps), \ \ \ \ \ \  \ \ \ \ \\\nonumber\\
\left(\begin{array}{cc}Q&0\\0&-Q'\end{array}\right) &\leq& \frac{3}{\eps} \left(\begin{array}{cc}I_2&-I_2\\-I_2&I_2\end{array}\right), \label{ineqMat}\\\nonumber \\
r+r'&=& 0. \label{q+q'}
\enq
Since $X_\eps$ converges to $\bar X$, we have $X_\eps$ $>$ $0$ for $\eps$ small enough, and we can use the viscosity subsolution property of $\hat{w}_1$ to obtain
\beq
-r + (\beta+\lambda) \hat{w}_1(t_\eps,X_\eps) - \widetilde{U}\left( \frac{x_\eps - x'_\eps}{\eps}\right)- \lambda G[\Hc \hat{w}_1](t_\eps,x_\eps,y_\eps) \;\;\;\;\;   &&  \label{ineqw1} \\
 - \sup_{\pi \in \R}\left[\pi \frac{x_\eps - x'_\eps}{\eps} + \frac{\rho b_L \sigma_I}{\sigma_L}y_\eps \frac{y_\eps - y'_\eps}{\eps} + \frac{\sigma_L^2 \pi^2}{2} Q_{11} +   \pi \rho \sigma_I \sigma_L y_\eps Q_{12} + \rho^2 \frac{\sigma_I^2}{2} y_\eps^2 Q_{22}\right]  &\leq& 0, \nonumber
\enq
and the supersolution property of $\hat{w}_{2,n}$ to get
\beq
-r' + (\beta+\lambda) \hat{w}_{2,n}(t_\eps,X'_\eps) - \widetilde{U}\left( \frac{x_\eps - x'_\eps}{\eps}\right)- \lambda G[\Hc \hat{w}_{2,n}](t_\eps,x'_\eps,y'_\eps) \;\;\;\;\;   &&  \label{ineqw2} \\
- \sup_{\pi \in \R}\left[\pi \frac{x_\eps - x'_\eps}{\eps} + \frac{\rho b_L \sigma_I}{\sigma_L}y'_\eps \frac{y_\eps - y'_\eps}{\eps}+\frac{\sigma_L^2 \pi^2}{2} Q'_{11} +   \pi \rho \sigma_I \sigma_L y'_\eps Q'_{12} + \rho^2 \frac{\sigma_I^2}{2} (y'_\eps)^2 Q'_{22}\right]  &\geq& 0 . \nonumber
\enq
Subtracting \reff{ineqw1} from \reff{ineqw2},  using the fact that the difference of the supremum is less than the supremum of the difference and \reff{q+q'}, we obtain
\beq
&&(\beta+\lambda) (\hat{w}_1(t_\eps,X_\eps) - \hat{w}_{2,n}(t_\eps,X'_\eps)) \nonumber \\
&\leq& \sup_{\pi \in \R} \left[ \frac{\sigma_L^2 \pi^2}{2} (Q_{11} - Q'_{11}) +   \pi \rho \sigma_I \sigma_L (y_\eps Q_{12} - y'_\eps Q'_{12}) + \rho^2 \frac{\sigma_I^2}{2} (y_\eps^2 Q_{22}-(y'_\eps)^2 Q'_{22})\right] \nonumber \\
&& \;\;\;+ \frac{\rho b_L \sigma_I}{\sigma_L}\frac{(y_\eps - y'_\eps)^2}{\eps} + \lambda \left( G[\Hc \hat{w}_1](t_\eps,X_\eps) - G[\Hc \hat{w}_{2,n}](t_\eps,X'_\eps)\right).  \label{ineqV}
\enq
Now notice that
\beq
\lim_{\eps \rightarrow 0} \left(G[\Hc \hat{w}_1](t_\eps,X_\eps) - G[\Hc \hat{w}_{2,n}](t_\eps,X'_\eps) \right)&=& G[\Hc \hat{w}_1](\bar t,\bar X) - G[\Hc \hat{w}_{2,n}](\bar t, \bar X) \nonumber \\
&\leq& \sup_{\R_+} (\Hc \hat{w}_1 - \Hc \hat{w}_{2,n}) \nonumber \\
&\leq& M. \label{ineqG}
\enq
Furthermore, using \reff{ineqMat} we see that  for all $\pi$ $\in$ $\R$
\beq
&& \frac{\sigma_L^2 \pi^2}{2} (Q_{11} - Q'_{11}) +   \pi \rho \sigma_I \sigma_L (y_\eps Q_{12} - y'_\eps Q'_{12}) + \rho^2 \frac{\sigma_I^2}{2} (y_\eps^2 Q_{22}-(y'_\eps)^2 Q'_{22}) \nonumber \\
 &&\nonumber \\
&=& \frac{1}{2} \left(\begin{array}{cccc} \sigma_L \pi &\rho \sigma_I y_\eps &  \sigma_L \pi &\rho \sigma_I y'_\eps \end{array} \right) \left(\begin{array}{cc}Q&0\\0&-Q'\end{array}\right)\left(\begin{array}{c}\sigma_L \pi \\\rho \sigma_I  y_\eps \\ \sigma_L\pi \\\rho \sigma_I y'_\eps \end{array} \right)\nonumber \\
 &&\nonumber \\
&\leq& \frac{1}{2}\left(\begin{array}{cccc}\sigma_L \pi & \rho \sigma_I y_\eps &\sigma_L \pi & \rho \sigma_I y'_\eps \end{array} \right) \left(\begin{array}{cc}I_2&-I_2\\-I_2&I_2\end{array}\right)\left(\begin{array}{c}\sigma_L \pi \\ \rho \sigma_I y_\eps \\\sigma_L \pi \\\rho \sigma_I y'_\eps \end{array} \right)\nonumber \\
 &&\nonumber \\
&\leq& (\rho \sigma_I)^2 \frac{3}{2\eps} |y_\eps - y'_\eps|^2 . \label{ineqA}
\enq
Recall that by \reff{limX-X'}
\beq
\frac{(y_\eps - y'_\eps)^2}{\eps} &\rightarrow& 0 \;\;\;\;\; \mbox{ when } \eps \rightarrow 0. \label{ineqY}
\enq
Letting $\eps$ go to $0$ in \reff{ineqV}, and combining \reff{ineqG}-\reff{ineqA}-\reff{ineqY}, we finally obtain
\beqs
(\beta+\lambda) M &\leq& \lambda M,
\enqs
so $M$ $\leq$ $0$.
\ep
\section{An iterative approximation scheme for the value functions}
In this section we present an iterative scheme to compute numerical approximations of  the value functions $V$ and $\widehat{V}$.
For sake of brevity we omit the proofs of the results that can be found in \cite{gasphd}.

First of all, we observe that \reff{HJBv} contains a nonlocal term, i.e. $G[\Hc \widehat{V}]$. Thus, in order to get a computational tool to approximate  $V$ and $\widehat{V}$,  it is needed  to couple standard numerical schemes with an iterative procedure as we are going to describe.

We start with 
\beq V^0 &=&0.  \label{V0}
\enq
Then,  inductively:  
 
\begin{itemize}
\item[-] Given $n\geq 0$ and $V^n$, we define $\widehat{V}^{n}$ on $\R_+^3$ as the unique (constrained viscosity) solution to
\beq \label{HJBvn}
- \widehat{V}^{n}_t  + (\beta+\lambda) \widehat{V}^{n}  - \lambda G[V^n]- \sup_{c\geq 0, \pi\in\R} H_{cv}(y,D_{(x,y)}\widehat{V}^{n},D^2_{(x,y)}, \widehat{V}^{n};c,\pi) \
 = \ 0,
\enq
with boundary condition
\beq \label{Bndryvn}
\widehat{V}^{n}(t,0,y)& =& \E \int_t^\infty e^{-(\beta+\lambda)(s-t)} \lambda G[V^{n}](s,0,\tilde{Y}^{t,y}_s) ds.
\enq
and growth condition
\beq \label{growthvn}
|\widehat{V}^{n}(t,x,y)| &\leq& K e^{\tilde{k}_p t} (x+y)^p.
\enq
\item[-]Given $n\geq 0$ and $\widehat V^{n}$, we define $V^{n+1}$ by
\beq \label{relVhatVn}
V^{n+1} &=& \Hc \widehat{V}^{n}.
\enq
\end{itemize}
We have a stochastic control representation for $(\widehat{V}^{n}, V^{n})_{n\geq 0}$:

\begin{proposition}
For each $n\geq 0$ we have 
\beq \label{defVn}
V^n(r) &=& \sup_{(c,\pi,\alpha) \in \Ac(r)} \E \int_0^{\tau_n} e^{- \beta t} U(c_s) ds,
\enq
and
\beq \label{defhatVn}
\widehat V^{n}(t,x,y) \;=\; \sup_{(c,\pi) \in \Ac_t(x)} \E\int_t^\infty e^{- (\beta+\lambda) (s-t)} \left( U({c}_s) + \lambda G[V^n]\left(s, \tilde{X}_s^{t,x,{\pi},{c}}, \tilde{Y}_s^{t,y}\right) \right)ds.
\enq
\end{proposition}
\medskip
%
%
%
The next result states the convergence of  $V^n$  to $V$ at an exponential rate.
\begin{proposition} \label{propVn}
For some $K>0$, we have
\beq
0 \ \ \leq \ \ (V - V^n)(r) &\leq& K r^p \delta^n,  \label{convVn} \\
0 \ \ \leq \  \ (\widehat V-\widehat V^{n})(t,x,y) &\leq& K e^{\tilde{k}_p t} (x+y)^p \delta^n, \label{convhatVn}
\enq
where
\beqs
\delta &:=& \frac{\lambda}{\lambda + \beta - k_p} \;\;<\;\; 1.
\enqs
\end{proposition}
%
%
\medskip
To solve the PDE \reff{HJBvn} one needs to approximate it by a finite horizon PDE. To this end, we fix some finite horizon $T>0$ and  consider the functions $\widehat{V}^{n,T}$, $V^{n,T}$ defined recursively as follows:

\begin{itemize}
\item[-] $V^{0,T} = 0$.
\item[-] Given $n\geq 0$ and $V^{n,T}$, and given some  terminal boundary condition condition $\phi^{n,T}$, we define on $[0,T] \times \R_+^2$
\beqs
\widehat{V}^{n,T}(t,x,y) &=&\sup_{(c,\pi) \in \Ac_t(x)} \E\left[\int_t^T e^{- (\beta+\lambda) (s-t)} \left( U({c}_s) + \lambda G[V^{n,T}]\left(s, \tilde{X}_s^{t,x,{\pi},{c}}, \tilde{Y}_s^{t,y}\right) \right)ds  \right.\\
&& \;\;\;\;\;\;\;\;\;\;\;\;\;\;\;\;\;\;\;\; \left.+ e^{-(\beta + \lambda)(T-t)} \phi^{n,T}(\tilde{X}_s^{t,x,{\pi},{c}}, \tilde{Y}_s^{t,y}) \right],
\enqs
\item[-] Given $n\geq 0$ and $\widehat{V}^{n,T}$ we define
\beqs \label{defVnT}
V^{n+1,T} &=& \Hc \widehat{V}^{n,T}
\enqs
\end{itemize}
By the same methods as above it is then straightforward to check that, for each $n\geq 0$, $\widehat{V}^{n,T}$ is a constrained viscosity solution on $[0,T) \times \R_+^2$ to
\beqs \label{eqHJBvnT}
- \widehat{V}^{n,T}_t  + (\beta+\lambda) \widehat{V}^{n,T}  - \lambda G[V^{n,T}](t,x,y)-\!\!\! \sup_{c\geq 0, \pi\in\R} H_{cv}(y,D_{(x,y)}\widehat{V}^{n,T},D^2_{(x,y)} \widehat{V}^{n,T};c,\pi) 
 \ =\ 0,
\enqs
 with boundary conditions
\beqs
\widehat{V}^{n,T}(T,x,y) &=& \phi^{n,T}(x,y), \\
\widehat{V}^{n,T}(t,0,y) &=& \E \left[\int_t^T e^{-(\beta+\lambda)(s-t)} \lambda G[V^{n,T}](s,0,\tilde{Y}^{t,y}_s) ds + e^{-(\beta + \lambda)(T-t)} \phi^{n,T}(0,\tilde{Y}_s^{t,x,{\pi},{c}})\right].
\enqs
Now we assume that the terminal condition $\phi^{n,T}$ satisfies 
\beqs \label{ineqTerm}
\left|\phi^{n,T}(x,y) - \widehat{V}^{n}(T,x,y) \right| &\leq& \Ec e^{\tilde{k}_p T} (x+y)^p,
\enqs
for some error $\Ec$ not depending on $n$.
Note that this assumption is not restrictive since \\ $0 \leq \widehat{V}^{n} \leq \widehat{V}$, and so due to \reff{Growthvhat}, the inequality \reff{ineqTerm} is satisfied, e.g., by taking $\phi^{n,T}=0$. 
\smallskip

We then have the following estimate for the numerical error induced by the finite horizon approximation:

\begin{proposition} \label{propT} 
For every $n\geq 1$ and every  $t$ $\in$ $[0,T]$,  $r$, $x$, $y$ $\in$ $\R_+$,
\beqs
 |(V^{n,T} - V^n)(r)| &\leq& \frac{\Ec}{1-\delta}\, e^{-(\beta+ \lambda -k_p) T} r^p, \\
 |(\widehat{V}^{n,T} - \widehat{V}^n)(t,x,y)| &\leq& \frac{\Ec}{1-\delta}\, e^{-(\beta+ \lambda -k_p) T} e^{\tilde{k}_p t} (x+y)^p.
\enqs
\end{proposition}
\medskip
By combining Propositions \ref{propVn} and \ref{propT}, one can choose $n$, $T$ large enough to approximate $V$, $\widehat{V}$ by  $V^{n,T}$,  $\widehat{V}^{n,T}$ (respectively) with any required precision.  The latter ones can be computed by the iterative procedure described above, using at each step of the iteration a standard explicit finite-difference scheme for parabolic viscosity solutions to solve the PDE (see e.g. chapter IX in \cite{FS} for a description of the scheme, as well as the proof of its convergence). 
Finally, we observe that the choice of $n$ and $T$ has to  depend on $\lambda$:
\begin{itemize}
	\item[-] When $\lambda$ is large, $\delta$ is close to $1$ so that the number of iterations $n$ must be chosen large.
	\item[-] The finite horizon error is roughly speaking of order $(1+\lambda) e^{- (1+ \lambda) T}$, so that $T$ may be chosen small for large $\lambda$ and must be reasonably large for small $\lambda$.
\end{itemize}

\section{Cost of illiquidity and optimal policy in the illiquid asset}\label{sec:optpol}
The results obtained allow  us to measure the cost of illiquidity and to determine the optimal policy allocation in the illiquid asset. Indeed,
$\widehat V$ can be computed numerically following the scheme described in Section 4, and then the optimal allocation $(\alpha^*_k)_{k\geq 0}$ in the illiquid asset and the value function $V$  can be derived.

At  $\tau_0=0$ the optimal allocation in the illiquid asset is
\beqs\label{DPPdiscrete0}
\alpha_0^*&=&\mbox{argmax}_{0\leq a\leq r}\ \widehat{V}(0,r-a, a).
\enqs
and consequently the value function $V$ can be computed.

\begin{figure}[h]
\include{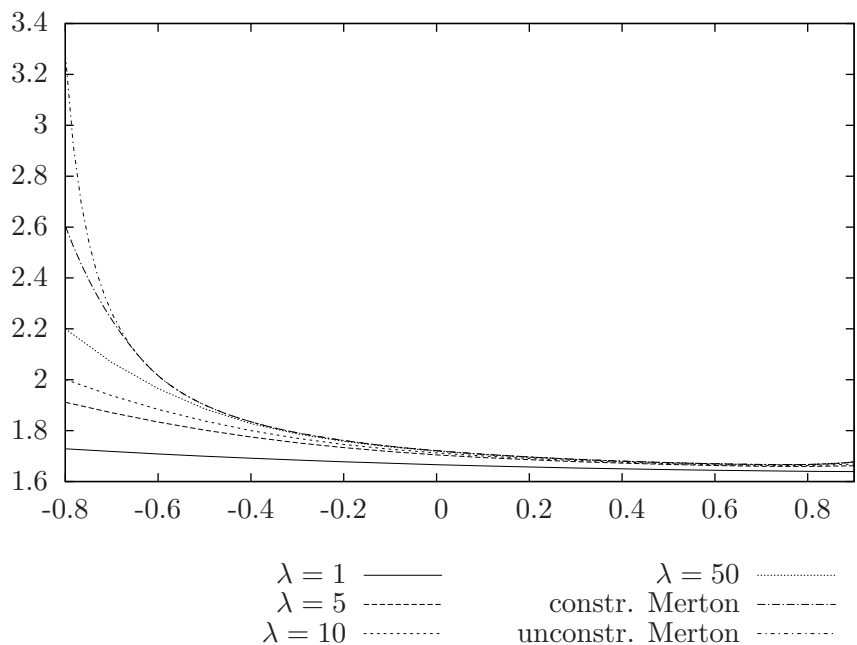}
\caption{{\small{Value function $V(1)$ as a function of $\rho$, for various $\lambda$.}}}
\label{figV}
\end{figure}

Figure \ref{figV}  shows the impact of illiquidity in the case of power utility $U(c)=c^p/p$.  In this case, by standard arguments using the homogeneity of $U$, one can prove that the value function has the structure $V(r)=V(1)r^p$. The value $V(1)$ is represented in Figure \ref{figV} as function of $\rho$ for different values of the liquidity parameter $\lambda$. The lines corresponding to the constrained and unconstrained  Merton refer to the problem when the asset $I$ is considered as liquid and when,  respectively, the constraint $\pi_I\in [0,1]$ is imposed or not.   The parameters are set as follows:
$$
\beta = 0.2, \;\;\;\;\;\;p=0.5,\;\;\;\;\;\;b_L = 0.15, \;\;\;\;\;\;\sigma_L = 1, \;\;\;\;\;\; b_I=0.2, \;\;\;\;\;\;\sigma_I = 1.
$$
We observe, as expected, a monotone convergence to the constrained Merton problem (see also \cite{FGGfin} for comments). The difference between the different values of $\lambda$ can be taken as an absolute measure of the cost of illiquidity.  

\begin{figure}[h]
\include{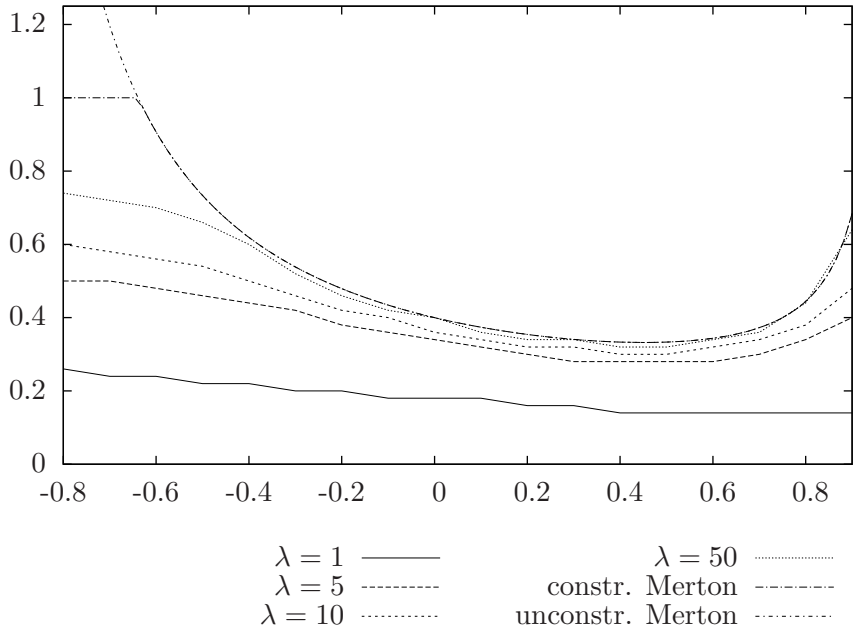}
\vspace{-1cm}
\caption{\footnotesize{Optimal investment proportion $\alpha_0^*/r$  in the illiquid asset as function of $\rho$ for various $\lambda$}.}
\label{figA}
\end{figure}

In Figure \ref{figA} we plot the optimal investment proportion in the illiquid asset $\alpha_0^*/r$ as a function of the correlation $\rho$, for various values of the liquidity parameter $\lambda$. Also in this case we observe the monotone convergence to the constrained Merton problem.
%
%
%
\medskip\\
\ni
{\small{\textbf{Acknowledgements.} The authors are particularly grateful to Huy\^en Pham to have suggested  the topic and to Fausto Gozzi for valuable suggestions on the presentation of the paper.
}}

\end{document}